\documentclass[final,5p,times,twocolumn]{elsarticle}
\usepackage{amssymb}
\usepackage[latin1]{inputenc}
\usepackage{amsmath}
\usepackage{graphicx}
\usepackage{epstopdf}
\usepackage{color}
\biboptions{sort&compress}
\journal{Chemical Physics}

\begin{document}
\begin{frontmatter}
\title{Nonclassical phase-space trajectories for the damped
harmonic quantum oscillator}

\address[label1]{Departamento de F\'{\i}sica, Universidad Nacional
	de Colombia, Bogot\'a D.C., Colombia.}

\address[label2]{Institut f\"ur Physik, Universit\"at Augsburg,
	Universit\"atsstra{\ss}e~1, D-86135 Augsburg, Germany}

\address[label3]{CeiBA -- Complejidad, Bogot\'a D.C., Colombia.}

\author[label1,label2,label3]{L.~A.~Pach\'on}

\author[label2]{G.-L.~Ingold}

\author[label1,label3]{T.~Dittrich}

\begin{abstract}
The phase-space path-integral approach to the damped harmonic oscillator is
analyzed beyond the Markovian approximation. It is found that pairs of
nonclassical trajectories contribute to the path-integral representation of the
Wigner propagating function. Due to the linearity of the problem, the sum
coordinate of a pair still satisfies the classical equation of motion.
Furthermore, it is shown that the broadening of the Wigner propagating
function of the damped oscillator arises due to the time-nonlocal
interaction mediated  by the heat bath.
\end{abstract}

\end{frontmatter}

%%%%%%%%%%%%%%%%%%%%%%%%%%%%%%%%%%%%%%%%%%%%%%%%%%%%%%%%%%%%%%%%%%%%%%%%%%%%%%
\section{Introduction}\label{Introduction}
%%%%%%%%%%%%%%%%%%%%%%%%%%%%%%%%%%%%%%%%%%%%%%%%%%%%%%%%%%%%%%%%%%%%%%%%%%%%%%
In quantum chemistry, biophysics, and more recently in some problems in quantum
computation, the time evolution of systems of large size and at high
excitations is in the center of interest. Generally, a full quantum treatment
of such systems is practically impossible. In order to get a first insight into
the behavior of such systems, they are usually propagated using the associated
classical equations of motion.  This approach, which disregards all possible
quantum effects, is known as molecular dynamics (MD). Often, the scales
involved in a problem call for a semiclassical approximation where molecular
dynamics would constitute the zeroth order. To next order,
one obtains the well-known van Vleck-Gutzwiller propagator
\cite{vV28,gut67}. This quantity is arguably one of the most
fundamental elements of semiclassical theory. It constitutes,
e.g., the starting point for the derivation of the celebrated Gutzwiller trace
formula \citep{gut71} and also for the semiclassical reaction rate
theory \cite{HTB90,HH91}.

Traditionally, semiclassical methods have been developed in position
representation. This implies, however, that i) determinants arising from the
projection along momentum appear in the prefactor, ii) they
are formulated in terms of double-sided boundary condition problems
and iii) are based on wave functions. Thus, they do not speak
the same language as classical mechanics, which certainly would be
desirable in order to describe and study the
quantum-classical transition \cite{ber89,TAO01,DP09}. In view of these
drawbacks, semiclassical methods in phase space offer a conceptually clearer
and possibly a numerically more efficient approach \cite{RO02,DVS06,DGP09}.

Recently, significant progress has been achieved as to semiclassical
approximations to the Wigner propagator corresponding to the unitary quantum
dynamical group. The case of damped nonunitary time evolution, giving rise to a
semi-group, is different even in its basic mathematical structures and largely
requires to reconsider the phase-space approach. In order to clearly
distinguish the two cases, we shall use the term ``propagating function''
instead of ``propagator'' in the context of this semi-group.

To be sure, a semiclassical phase-space analysis of open quantum systems has
recently been presented in the Markovian limit \cite{ORB09} on the basis of the
Lindblad master equation which can be extended to the general non-Markovian
case \cite{PDI10} for a system bilinearly coupled to a bath consisting of
harmonic oscillators.  In order to illustrate the latter approach by means of
an analytically solvable case, we here present a study of the damped harmonic
oscillator in phase space.  While certain aspects will turn out to be specific
for this linear system, we expect the overall picture
to be useful to understand the more general case of nonlinear
systems for which details will be discussed elsewhere \cite{PDI10}.

In an attempt to make the discussion accessible to readers with different
backgrounds, we review in the next three sections the basic ideas of the
semiclassical phase-space approach, the influence functional theory, and the
quantum damped harmonic oscillator as far as they are needed for the following.
Section~\ref{sec:qdhoPhaseSpace} will then contain the main results for the
propagating function in phase space.  We will argue that a Gaussian broadening
arises due to the nonlocal interaction of the system degree of freedom mediated
by the coupling to the environment.  Furthermore, we demonstrate the
nonclassical nature of the trajectories contributing within the phase-space
path-integral representation of the Wigner propagating function.

%%%%%%%%%%%%%%%%%%%%%%%%%%%%%%%%%%%%%%%%%%%%%%%%%%%%%%%%%%%%%%%%%%%%%%%%%%%%%%
\section{Unitary evolution of the density matrix}\label{sec:unitary}
%%%%%%%%%%%%%%%%%%%%%%%%%%%%%%%%%%%%%%%%%%%%%%%%%%%%%%%%%%%%%%%%%%%%%%%%%%%%%%
Before discussing the case of a damped quantum system, it is useful to fix the
main ideas and the notation by considering the unitary time evolution of an
isolated system degree of freedom in real space as well as in phase space. In
later sections, we will be forced to employ a density operator to describe the
state of a damped quantum system. Therefore, we express the state of our
isolated system S, which could well be a pure state, also in terms of a density
matrix denoted by $\hat\rho_\mathrm{S}$ in the following.

For an isolated system described by a time-independent Hamiltonian $\hat
H_\mathrm{S}$, the time evolution of an initial density matrix
$\hat\rho_\mathrm{S}(0)$ is determined through the unitary time evolution
operator
\begin{equation}
\label{equ:TimeEvoOp}
\hat U(t) = \exp\left(-\frac{{\rm i}}{\hbar} {\hat H}_{\rm S} t\right)
\end{equation}
and its adjoint operator $\hat U^\dagger(t)$ by means of the relation
\begin{equation}
\label{equ:DefEvoRHO}
{\hat \rho}_\mathrm{S}(t) = {\hat U}(t) {\hat \rho}_{\rm S}(0) {\hat U}^{\dag}(t).
\end{equation}
In position representation, this expression turns into
\begin{equation}
\label{equ:rhoRP}
\rho_\mathrm{S}(q_+'',q_-'',t) =
\int\mathrm{d} q_+' \mathrm{d} q_-'  J(q_+'',q_-'',t; q_+',q_-',0)
\rho_\mathrm{S}(q_+',q_-',0) \,,
\end{equation}
where
\begin{equation}
\label{equ:PropDensMatUnit}
J(q_+ '',q_- '',t; q_+ ',q_- ',0)= U(q_+'',q_+',t)U^*(q_-'',q_-',t)
\end{equation}
is the propagator with
\begin{equation}
U(q_\pm'',q_\pm',t) = \langle q_\pm'' \vert\hat U(t)\vert q_\pm'\rangle
\end{equation}
and
\begin{equation}
\rho_\mathrm{S}(q_+,q_-) = \langle q_+\vert\hat\rho_\mathrm{S}\vert q_-\rangle
\end{equation}
is the density matrix.

According to Feynman \cite{fey48,FH65}, the unitary propagator can be expressed
as a path integral
\begin{equation}
\label{equ:DefPropPI}
U(q_{\pm}'',q_{\pm}',t) =
\int {\cal D} q_{\pm} \exp\left[\frac{{\rm i}}{\hbar}
S_{\rm S}(q_{\pm},t)\right]
\end{equation}
running over all paths which satisfy the boundary conditions
$q_{\pm}(0)=q_{\pm}'$ and $q_{\pm}(t)=q_{\pm}''$.
\begin{equation}
S_{\rm S}[q,t]=\int_{0}^{t} \mathrm{d}t' L_{\rm S} (q,\dot q,t')
\end{equation}
is the action associated with the path $q(t')$ where $L_{\rm S}(q,\dot
q,t')$ is the Lagrangian describing degree of freedom of the system.

As a system degree of freedom, we will specifically consider a harmonic
oscillator of mass $m$ and frequency $\omega_0$. Then, the exact propagator
takes the form
\begin{equation}
\label{equ:feypropgHO}
U(q_{\pm}'', q_{\pm}', t) =
\left(\frac{1}{2\pi\mathrm{i}\hbar}
\frac{\partial^2 S_{\rm S}^\mathrm{cl}}{\partial q_{\pm}''
\partial q_{\pm}'}\right)^{\frac{1}{2}}
\exp \left[ \frac{\mathrm{i}}{\hbar}
S_\mathrm{S}^\mathrm{cl}(q_{\pm}'',q_{\pm}',t) \right]\,.
\end{equation}
The classical action
\begin{equation}
S_\mathrm{S}^\mathrm{cl} (q_\pm'',q_\pm',t) =
\frac{m\omega_0}{2\sin(\omega_0 t)}
\left[(q_{\pm}''^2 + q_{\pm}'^2 )
\cos\left(\omega_0 t \right) - 2 q_{\pm}'' q_{\pm}'\right]
\end{equation}
is completely determined by solutions of the classical equation of motion
\begin{equation}
\label{equ:TrajUnitEvol}
\ddot{q}_{\pm}^{\rm cl} = - m \omega_0^2 q_{\pm}^{\rm cl}\,,
\end{equation}
while the fluctuations contribute only to the prefactor which is independent of
$q_\pm''$ and $q_\pm'$.

For the harmonic oscillator in real space, we can thus conclude that the
propagator for the density matrix is determined by two independent solutions of
the classical equation of motion. The question now arises what can be said
about the relevant trajectories in phase space where instead of two initial and
two final points in position space one has only one initial and one final point
in phase space, each comprising a position and a momentum.

Among the infinitely many possible phase-space representations
\cite{HOSW84} we choose the Wigner function, which for our
one-dimensional system degree of freedom is given by the Weyl-ordered
transform of the density operator \cite{HOSW84},
\begin{equation}\label{equ:wigfunc}
W_{\rm S}({\bf r}) =  \int\frac{\mathrm{d}q'}{2\pi\hbar}
\exp\left(-\frac{\mathrm{i}}{\hbar}p q'\right)
\left\langle q+\frac{q'}{2} \left|
\hat\rho_\mathrm{S}\vphantom{\frac{q'}{2}}\right|
q-\frac{q'}{2}\right\rangle\,.
\end{equation}
Here, ${\bf r} = ({p},{q})$ denotes a vector in two-dimensional phase space.

\begin{figure}
\centerline{\includegraphics[width=0.6\columnwidth]{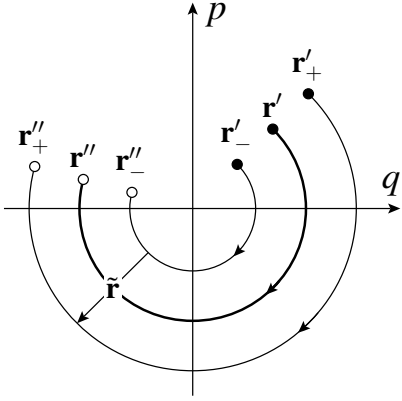}}
\caption{For an undamped harmonic oscillator, the pairs of phase-space
trajectories $\mathbf{r}_{\pm}$ as well as their sum trajectory $\mathbf{r}$
satisfy the classical equations of motion.  $\mathbf{r}'_{\pm}, \mathbf{r}'$
and $\mathbf{r}''_{\pm}, \mathbf{r}''$ denote the initial and final
phase-space points of the trajectories, respectively, and
$\tilde{\mathbf{r}}=\mathbf{r}_{+}-\mathbf{r}_{-}$.}
\label{fig:WignerPropUni}
\end{figure}

By means of the transformation (\ref{equ:wigfunc}) one obtains from
(\ref{equ:rhoRP}) for the time evolution of the Wigner function
\begin{equation}\label{equ:wigprop}
W_{\rm S}({\bf r}'',t) = \int {\rm d}^{2}r'\,G_{\rm W}({\bf r}'',{\bf r}',t)
W_{\rm S}({\bf r}',0)\,.
\end{equation}
Introducing the difference coordinate $\tilde q = q_+ - q_-$ and the sum
coordinate $q = (q_+ + q_-)/2$, the Wigner propagator appearing here as an
integral kernel can be written in terms of a double Fourier transform of the
propagator of the density matrix (\ref{equ:PropDensMatUnit}) along the
difference coordinates
\begin{equation}
\label{equ:DefUUWignerProp}
\begin{split}
G_\mathrm{W}(\mathbf{r}'',\mathbf{r}',t) = \frac{1}{2\pi\hbar}
\int{\rm d}\tilde q'{\rm d}\tilde q''
\exp\left[\frac{{\rm i}}{\hbar} (p'\tilde q' - p''\tilde q'' ) \right] \\
\times U\left(q''+\frac{\tilde q''}{2},q'+\frac{\tilde q'}{2},t \right]
U^*\left(q''-\frac{\tilde q''}{2},q'-\frac{\tilde q'}{2},t \right)\,.
\end{split}
\end{equation}
A similar expression can be obtained with the position-space
propagator replaced by the Weyl transform of the time evolution
operator \cite{mar91},
\begin{equation}
\label{equ:DefWeylWignerProp}
\begin{split}
G_\mathrm{W}(\mathbf{r}'',\mathbf{r}',t) = \frac{1}{2\pi\hbar}
\int{\rm d}^2\tilde r
\exp\left[\frac{\rm i}{\hbar} (\mathbf{r}' - \mathbf{r}'')
\wedge\tilde{\mathbf{r}} \right] \\
\times U_{\rm W}\left(\frac{\tilde{\mathbf{r}}'+\tilde{\mathbf{r}}''}{2}+
\frac{\tilde{\mathbf{r}}}{2},t\right)
U_{\rm W}^*\left(\frac{\tilde{\mathbf{r}}'+\tilde{\mathbf{r}}''}{2}-
\frac{\tilde{\mathbf{r}}}{2},t\right)\, ,
\end{split}
\end{equation}
where, analogous to (\ref{equ:wigfunc}),
\begin{equation}\label{equ:weylprop}
U_{\rm W}({\bf r},t) =  \int\frac{\mathrm{d}q'}{2\pi\hbar}
\exp\left(-\frac{\mathrm{i}}{\hbar}p q'\right)
\left\langle q+\frac{q'}{2} \left|
\hat U(t)\vphantom{\frac{q'}{2}}\right|
q-\frac{q'}{2}\right\rangle\,.
\end{equation}
In this way, the classical trajectories $q_\pm^\mathrm{cl}$
determining the propagator (\ref{equ:feypropgHO}) in position space
become the phase-space trajectories $\mathbf{r}_\pm$ shown in
Fig.~\ref{fig:WignerPropUni}. All such pairs share the same mean
$\mathbf{r} = (\mathbf{r}_- + \mathbf{r}_+)/2$. For the harmonic
oscillator, the phase-space difference vector $\tilde{\mathbf{r}} =
\mathbf{r}_+ - \mathbf{r}_-$ appears only linearly in the dynamical
phase. Thus, integration over $\tilde{\mathbf{r}}$ in
(\ref{equ:DefWeylWignerProp}) directly results in the exact Wigner
propagator
\begin{align}
\label{equ:wigpropgHODelta}
G_\mathrm{W}({\bf r}'',{\bf r}',t) =
\delta\left[ {\bf r}'' - {\bf r}^{\rm cl}({\bf r}',t)\right]\,.
\end{align}
This result implies that each initial point $\mathbf{r}'$ in phase space is
propagated along the solution $\mathbf{r}^\mathrm{cl}(\mathbf{r}',t)$ of the
classical equation of motion. All quantum effects, including tunneling
\cite{BV90}, must therefore be contained already in nonclassical features, 
such as an irreducible energy spread, of the initial Wigner function itself.

As we shall see in Sect.~\ref{sec:qdhoPhaseSpace}, the
relevant trajectories will be different for the harmonic oscillator in
the presence of damping. While the mean of a trajectory pair
still satisfies the classical equation of motion, the two trajectories
themselves will behave nonclassically and separate exponentially fast
in time. How this comes about will be explained in the following sections.

%%%%%%%%%%%%%%%%%%%%%%%%%%%%%%%%%%%%%%%%%%%%%%%%%%%%%%%%%%%%%%%%%%%%%%%%%%%%%%
\section{Influence functional theory}\label{sec:influenceFunctional}
%%%%%%%%%%%%%%%%%%%%%%%%%%%%%%%%%%%%%%%%%%%%%%%%%%%%%%%%%%%%%%%%%%%%%%%%%%%%%%
We now turn the so far isolated quantum system S into a dissipative quantum
system by coupling it to a heat bath characterized by a temperature $T$. The
complete Hamiltonian then is of the general form
\begin{equation}
\label{equ:TotalHamiltonian}
{\hat H} =  {\hat H}_\mathrm{S}+ {\hat H}_\mathrm{B} +  {\hat H}_\mathrm{SB}\,,
\end{equation}
where the second and third term describe the bath Hamiltonian and the
system-bath coupling, respectively.

As long as the complete Hilbert space is retained, the evolution of the density
matrix is still of the form (\ref{equ:DefEvoRHO}) where the system density
operator $\hat\rho_\mathrm{S}$ is replaced by the full density operator
$\hat\rho$. Correspondingly, in the time evolution operator
(\ref{equ:TimeEvoOp}) the system Hamiltonian has to be replaced by the full
Hamiltonian (\ref{equ:TotalHamiltonian}). In position representation,
(\ref{equ:rhoRP}) still holds with the appropriate replacements. In particular
the coordinates are now $\mathcal{Q} = \{q,\mathbf{Q}\}$ and comprise the
system coordinate $q$ as well as the vector of bath coordinates $\mathbf{Q}$.

If one considers a harmonic oscillator bilinearly coupled to a bath of
oscillators \cite{U66}, a situation which we will discuss in detail below, the
Wigner propagator
\begin{equation}
G_{\rm W}({\bf \mathcal{R}}'',{\bf \mathcal{R}}',t) =
\delta\left[{\bf \mathcal{R}}'' -
{\bf \mathcal{R}}^{\rm cl}({\bf \mathcal{R}}',t)\right].
\end{equation}
can be exactly expressed in terms of the classical phase-space trajectories
$\mathcal{R} = \{\mathbf{r}, \mathbf{R}\}$ of system and heat
bath. They are obtained even in the continuum limit and for a thermal 
initial state of the bath by transforming to normal modes of the total 
system \cite{PSZ08}. It is only after tracing out the heat bath that 
quantum effects come into play in the time evolution of the Wigner 
function. Instead of carrying out this analysis in phase space, we will 
rather follow the strategy presented in Sect.~\ref{sec:unitary} for the 
undamped case where we started from the position-space description.

For a dissipative system, one is usually not interested in the full dynamics
but only in the reduced dynamics of the system degree of freedom. In order to
integrate out the bath degrees of freedom, one needs to specify the initial
density matrix of system and bath. For simplicity, in the main part of the
paper we will restrict ourselves to factorizing initial conditions
\cite{FV63,CL83}. Then, the initial density matrix
\begin{equation}
\label{equ:initialState}
\hat\rho(0) = \hat\rho_\mathrm{S}(0)\frac{\exp(-\beta\hat H_\mathrm{B})}{Z_\mathrm{B}}
\end{equation}
is given by the product of the initial density matrix $\hat\rho_\mathrm{S}(0)$
of the system and the thermal density matrix of the heat bath at inverse
temperature $\beta=1/k_\mathrm{B}T$. $Z_\mathrm{B}$ denotes the partition
function of the bath. As we will show in \ref{sec:AppendixA}, our reasoning
can readily be generalized to a certain class of nonfactorizing initial conditions.

The time evolution of the initial state (\ref{equ:initialState}) is obtained as
a generalization of the considerations in Sect.~\ref{sec:unitary} by
substituting the single system degree of freedom by the ensemble of system and
bath degrees of freedom. In order to obtain the reduced dynamics of the system,
one needs to trace out the environmental degrees of freedom. This can be done
analytically if the heat bath is modelled by a set of harmonic oscillators with
masses $m_j$ and frequencies $\omega_j$ whose coordinates are bilinearly
coupled to the system coordinate. The Hamiltonian (\ref{equ:TotalHamiltonian})
then consists of the three contributions \cite{wei93}
\begin{align}
 {\hat H}_\mathrm{S} &=
	\frac{{\hat p}^2}{2m} + V({\hat q})\,, \label{equ:TotHamilS}
\\
 {\hat H}_\mathrm{B} &=
	\sum_{j=1}^\infty \left[\frac{1}{2m_j}{{\hat P}_j}^2 +
\frac{1}{2} m_j \omega_j^2 {\hat Q}_j^2\right]\,,
\\
\label{equ:InteractionH}
 {\hat H}_\mathrm{SB} &=
	-{\hat q}  \sum_{j=1}^\infty c_j {\hat Q}_j +
	{\hat q}^2 \sum_{j=1}^\infty \frac{c_j^2}{2 m_j \omega_j^2}\,,
\end{align}
where $c_j$ are the coupling constants. The second term in
(\ref{equ:InteractionH}) corrects a potential renormalization induced by the
coupling of the system to the heat bath. It turns out that the microscopic
details of the heat bath and its coupling to the system appear in the reduced
system dynamics only through the spectral density of bath oscillators
\begin{equation}
I(\omega) = \pi\sum_{j=1}^\infty\frac{c_j^2}
{2m_j\omega_j}\delta(\omega-\omega_j)\,.
\end{equation}
For later convenience, we introduce two quantities which depend on this
spectral density. The friction kernel is defined as
\begin{equation}
\label{equ:frictionKernel}
\eta(t) = \frac{2}{\pi}\int_0^\infty\mathrm{d}\omega
\frac{I(\omega)}{\omega}\cos(\omega t)
\end{equation}
and the correlation function of the noise \cite{HI05} induced by the coupling
to the heat bath is given by
\begin{equation}
\label{equ:noiseCorr}
K(z) = \int_0^\infty\frac{\mathrm{d}\omega}{\pi}I(\omega)
\frac{\mathrm{cosh}[\omega(\frac{1}{2}
\hbar\beta-\mathrm{i}z)]}{\mathrm{sinh}(\hbar\beta\omega/2)}\,,
\end{equation}
where $z$ is a generally complex time.

Tracing out the heat bath, one finds for the time evolution of the reduced
density matrix of the system starting from the factorizing initial state
(\ref{equ:initialState}) \cite{FV63,CL83}
\begin{equation}
\label{rhoRPKRJ}
\rho_{\rm S} (q_+'',q_-'',t) = \int\! \mathrm{d} q_+'\,\mathrm{d}
q_-' J(q_-'',q_+'',t; q_+',q_-')\,\rho_\mathrm{S}(q_+', q_-',0)\,.
\end{equation}
Here we have introduced the propagating function which can be expressed in
terms of a path integral over the system degree of freedom as
\begin{equation}
\label{equ:DefPropFunct}
\begin{aligned}
   J(q_-'',q_+'',t; q_+',q_-') &= \frac{1}{Z}
   \int {\cal D}q_+{\cal D}q_-
\\
   &\hspace{-2truecm}\times\exp\left\{\frac{\mathrm{i}}{\hbar}
   [S_{\rm S}(q_+) - S_{\rm S}(q_-)]\right\}
   {\cal F}(q_+,q_-)\,.
\end{aligned}
\end{equation}
The partition function $Z$ appearing here is an effective partition function of
the damped system defined as the ratio of the partition functions of system
plus bath and of the heat bath alone. The influence of the heat bath in
(\ref{equ:DefPropFunct}) is contained in the influence functional
\begin{equation}
\label{equ:ExplInfFunct}
\mathcal{F}(q_+,q_-) = \exp\left[-\frac{{1}}{\hbar} \Phi(q_+,q_-)\right]
\end{equation}
with the exponent
\begin{equation}
\begin{aligned}
\Phi[q_+,q_-] &= \int_0^t\mathrm{d}s\int_0^s\mathrm{d}u\, K'(s-u)
                 \big[q_+(s)-q_-(s)\big] \\
	      &\hspace{0.3\columnwidth}\times\big[q_+(u)-q_-(u)\big] \\
	      &\quad+\frac{\mathrm{i}}{2}
               \int_0^t\mathrm{d}s\int_0^s\mathrm{d}u\,
	       \eta(s-u)\big[q_+(s)-q_-(s)\big] \\
	      &\hspace{0.3\columnwidth}
               \times\big[\dot q_+(u)+\dot q_-(u)\big] \\
	      &\quad+\frac{\mathrm{i}}{2}\big[q_+'+q_-'\big]
               \int_0^t\mathrm{d}s\,
	       \big[q_+(s)-q_-(s)\big]\,.
\end{aligned}
\end{equation}
The prime in the first line indicates that the real part of the noise correlation
function (\ref{equ:noiseCorr}) should be taken.

%%%%%%%%%%%%%%%%%%%%%%%%%%%%%%%%%%%%%%%%%%%%%%%%%%%%%%%%%%%%%%%%%%%%%%%%%%%%%%%
\section{Quantum damped harmonic oscillator}\label{sec:qdho}
%%%%%%%%%%%%%%%%%%%%%%%%%%%%%%%%%%%%%%%%%%%%%%%%%%%%%%%%%%%%%%%%%%%%%%%%%%%%%%%
While the results reviewed in the previous section are valid for a general
system degree of freedom, we will now specifically consider a damped harmonic
oscillator with
\begin{equation}
\label{equ:potentialHO}
V(\hat q) = \frac{m}{2}\omega_0^2\hat q^2\,.
\end{equation}
In this way, we will be able to generalize the considerations of
Sect.~\ref{sec:unitary} concerning the relevant trajectories in phase
space. At the same time, this prevents making direct use of
semiclassical phase-space dynamics based on the van-Vleck approximation 
as in Refs.~\cite{DVS06,DGP09}, since this only applies to
sufficiently anharmonic potentials.

As for the propagator in the undamped case, the path-integral
expression for the propagating function (\ref{equ:DefPropFunct}) is
evaluated by an expansion around the paths maximizing the complex
action. The dependence on the initial and final coordinates is
entirely determined by these paths while the fluctuations only yield a
time-dependent prefactor. For a harmonic oscillator, the complex
action in (\ref{equ:DefPropFunct}) is stationary for trajectories
satisfying
\begin{equation}
\label{equ:GenStationaryPathsReT}
\begin{aligned}
&m {\ddot q}_{\pm}(s) + m \omega_0^2  q_{\pm}(s)
 \mp \frac{1}{2}\frac{{\rm d}}{{\rm d}s} \int_{s}^{t} {\rm d}u \,
\eta(s-u)\big[q_+(u) - q_-(u)\big]
\\
&+ \frac{1}{2} \frac{\rm d}{{\rm d}s} \int_0^{s} {\rm d}u \,
\eta(s-u)\big[q_+(u) + q_-(u)\big]=
\\
& \hspace{0.2\columnwidth}{\rm i}
\int_0^t {\rm d}u K'(s-u)\big[q_+(u)-q_-(u)\big]\,.
\end{aligned}
\end{equation}
As in the undamped case, the paths are subject to the boundary conditions
$q_{\pm}(0) = q_{\pm}'$ and $q_{\pm}(t) = q_{\pm}''$.

The two equations of motion (\ref{equ:GenStationaryPathsReT}) replace the
equations of motion (\ref{equ:TrajUnitEvol}) obtained for the undamped case.
As a consequence of the coupling to the heat bath, the equations of motion
(\ref{equ:GenStationaryPathsReT}) now include damping terms depending on the
friction kernel (\ref{equ:frictionKernel}). In addition, an imaginary nonlocal
force appears indicating the occurrence of decoherence. For linear systems, it
turns out that the imaginary part of the trajectories does not need to be
considered and that their real part is sufficient to obtain
the propagating function \cite{GSI88,HA85}.  As we shall see in the sequel,
neither of the two paths follows a classical equation of motion and their
separation grows exponentially fast. This somewhat surprising behavior is a
consequence of the coupling to the heat bath.

In order to make the discussion transparent, we now consider the
special case of Ohmic damping in addition to the assumption of factorizing
initial conditions and the restriction to the real part of the equations of
motion. We thus set $I(\omega)=m\gamma\omega$ and
$\gamma(t)=\eta(t)/m=2\gamma\delta(t)$, so that the equations of motion
(\ref{equ:GenStationaryPathsReT}) reduce to \cite{CL83}
\begin{equation}
\label{equ:GenStationaryPathsHO}
{\ddot q}_{\pm} + \omega_0^2 q_{\pm} + \gamma  {\dot q}_{\mp} = 0\,,
\end{equation}
where the damping term couples the trajectories $q_+$ and $q_-$.  It
is interesting to note that $\gamma q_{\mp}$ acts actually as a
driving instead of a damping in the sense that the separation between
trajectories grows exponentially (see
Fig.~\ref{fig:trajectories}). This can be seen more clearly by
decoupling the two equations of motion, using sum, $q = (q_+ + q_-)/2$, 
and difference coordinates, $\tilde{q} = q_+ - q_-$. 
The equations (\ref{equ:GenStationaryPathsHO}) then read
\begin{equation}
\label{equ:qxmotion}
\begin{aligned}
\ddot{q}   + \gamma \dot{q} +\omega_0^2 q &=0\\
\ddot{\tilde{q}} - \gamma \dot{\tilde{q}} + \omega_0^2 \tilde{q}&=0\,.
\end{aligned}
\end{equation}

As in the undamped case, the sum trajectory corresponding to the paths $q_+$
and $q_-$ obeys the classical equation of motion, which here takes a time-local
form because we have assumed Ohmic damping. In contrast to the sum coordinate
$q$ which decreases exponentially in time, the difference coordinate
$\tilde{q}$ grows exponentially so that we obtain a hyperbolic
dynamics in the $(q,\tilde{q})$-plane. As a consequence, the
trajectories $q_\pm$ do not obey the classical damped equation of
motion. The solutions of the equations of motion (\ref{equ:qxmotion}) read
\begin{equation}
\label{equ:ExplStationaryPathHO}
\begin{aligned}
q(s) &= q' \frac{G_-(t-s)}{G_-(t)} + q'' \frac{G_+(s)}{G_+(t)}\,,
\\
\tilde{q}(s) &= \tilde{q}' \frac{G_+(t-s)}{G_+(t)} +
\tilde{q}'' \frac{G_-(s)}{G_-(t)}\,,
\end{aligned}
\end{equation}
where
\begin{equation}
\label{equ:gpm}
G_{\pm}(t) = \frac{1}{\omega_\mathrm{d}}
\exp\left(\mp\frac{\gamma}{2}t\right)
\sin\left(\omega_\mathrm{d}t\right)
\end{equation}
and $\omega_\mathrm{d}^2 = \omega_0^2 - \gamma^2/4$. We remark that by choosing
appropriate functions $G(t)$, more general linear damped system like the
parametrically driven damped harmonic oscillator \cite{ZJH94,ZH95,KDH97} can be
studied with solutions of the form (\ref{equ:ExplStationaryPathHO}).

We now return to the propagating function which was given in
(\ref{equ:DefPropFunct}) in terms of a twofold path integral. It is
instructive to decompose the exponent into two parts
\begin{equation}
\label{equ:exponentPropFunc}
S(\tilde{q}'',q'',t;\tilde{q}',q') = S_1+S_2\,,
\end{equation}
where
\begin{equation}
\label{equ:exponent1}
S_1 = m\left[(q'\tilde{q}'+q''\tilde{q}'')\frac{\dot G_+(t)}{G_+(t)}
            -q'\tilde{q}''\frac{1}{G_-(t)}-q''\tilde{q}'\frac{1}{G_+(t)}\right]
\end{equation}
is obtained by evaluating the action of the system degree of freedom along the
trajectories given by (\ref{equ:ExplStationaryPathHO}) while
\begin{equation}
\label{equ:exponent2}
S_2 = \frac{\mathrm{i}}{2}\int_0^t {\rm d}s
\int_0^t {\rm d}u \, K'(s - u) \tilde{q}(s)\tilde{q}(u)
\end{equation}
arises from the influence functional (\ref{equ:ExplInfFunct}), i.e.\ by the
interaction of paths at different times through the coupling to the
environment. The significance of this decomposition will become clear in the
following section where we discuss the results of the present section from a
phase-space point of view.

%%%%%%%%%%%%%%%%%%%%%%%%%%%%%%%%%%%%%%%%%%%%%%%%%%%%%%%%%%%%%%%%%%%%%%%%%%%%%
\section{Quantum damped harmonic oscillator in phase space}
\label{sec:qdhoPhaseSpace}
%%%%%%%%%%%%%%%%%%%%%%%%%%%%%%%%%%%%%%%%%%%%%%%%%%%%%%%%%%%%%%%%%%%%%%%%%%%%%%

The undamped time evolution (\ref{equ:wigprop}) of the Wigner function can
immediately be transferred to the dissipative case if we relate the Wigner
propagating function to the propagating function by means of
\begin{equation}
\begin{aligned}
\label{equ:DefFVWignerProp}
G_\mathrm{W}(\mathbf{r}'',\mathbf{r}',t) &= \frac{1}{2\pi\hbar}
\int  {\rm d}\tilde{q}'{\rm d}\tilde{q}''
\exp\left[\frac{{\rm i}}{\hbar} (p'\tilde{q}' - p''\tilde{q}'') \right]
\\
&\hphantom{ \frac{1}{2\pi\hbar}\int}
\times J(\tilde{q}'',q'',t; \tilde{q}',q')\,,
\end{aligned}
\end{equation}
where we recall that $\mathbf{r}=(p,q)$ is the phase-space vector. Inserting
(\ref{equ:PropDensMatUnit}) valid for the unitary case into
(\ref{equ:DefFVWignerProp}) one recovers expression (\ref{equ:DefUUWignerProp})
for the Wigner propagator. We continue to restrict ourselves to factorizing
initial conditions but present in \ref{sec:AppendixA} the generalization of
(\ref{equ:DefFVWignerProp}) to nonfactorizing initial conditions.

Before analyzing the Wigner propagating function for Ohmic damping, we discuss
the decomposition (\ref{equ:exponentPropFunc}) of the exponent of the
propagating function. The contribution (\ref{equ:exponent1}) is linear in the
difference coordinates $\tilde{q}'$ and $\tilde{q}''$.  Performing the
transformation (\ref{equ:DefFVWignerProp}), we therefore arrive at the
Wigner propagating function
\begin{equation}
\label{equ:WPropDamped}
G_\mathrm{W}({\bf r}'',{\bf r}',t)= \delta\left[{\bf r}''-
{\bf r}^{\rm cl}({\bf r}',t)\right]\,,
\end{equation}
where the classical phase-space trajectory
\begin{equation}
\begin{aligned}
\label{classicalsolutionDHOps}
p^{\rm cl}(t) &= \dot G_+(t)p'+m\left\{\frac{[\dot G_+(t)]^2}{G_+(t)}-
\frac{1}{G_-(t)}\right\}q' \\
q^{\rm cl}(t) &= \frac{G_+(t)}{m}p' + \dot{G}_+(t)q',
\end{aligned}
\end{equation}
with $G_\pm(t)$ defined by (\ref{equ:gpm}) is now damped. While
(\ref{classicalsolutionDHOps}) satisfies $q^\mathrm{cl}(0)=q'$ as expected, the
initial momentum is given by $p^\mathrm{cl}(0)=p'-m\gamma q'$. This initial
slip is typical for factorizing initial conditions \cite{GSI88,HR85}.

Employing the Wigner propagating function (\ref{equ:WPropDamped}) amounts to
adding a velocity-dependent force in the system Hamiltonian as was proposed by
Caldirola and Kanai (see e.g. the review \cite{CYG02} for an
account of this kind of phenomenological approach). However, the
Wigner propagating function (\ref{equ:WPropDamped}) accounts only for
part of the exponent of the propagating function. The second
contribution (\ref{equ:exponent2}) is quadratic in the difference
coordinate and limits their contributions. As a result, the delta
function in (\ref{equ:WPropDamped}) will be broadened into a Gaussian
\begin{multline}
\label{WignerPropExplSimpl}
G_\mathrm{W}( {\bf r}'',t; {\bf r}') =
\frac{m}{2 \pi \hbar\Lambda(t)^{1/2}}
\left\vert\frac{\dot G_+(t)}{G_+(t)} \right\vert
\\
\times\exp\left\{- \frac{1}{2\hbar\Lambda(t)} \left[{\bf r}'' -
{\bf r}^{\rm cl}(t)\right]^{\rm T}
\mathsf{\Sigma} \left[{\bf r}'' - {\bf r}^{\rm cl}(t)\right] \right\},
\end{multline}
whose center moves along the damped classical trajectory
(\ref{classicalsolutionDHOps}).  The matrix appearing in the exponent is given
by its components
\begin{equation}
\begin{aligned}
\mathsf{\Sigma}_{11} &= a(t)\\
\mathsf{\Sigma}_{12} &= \mathsf{\Sigma}_{21} = -m\frac{\dot
G_+(t)}{G_+(t)}
\left[a(t)+b(t)\right]\\
\mathsf{\Sigma}_{22} &= m^2\frac{[\dot G_+(t)]^2}{[G_+(t)]^2}
\left[a(t)+2b(t)+c(t)\right]\\
\end{aligned}
\end{equation}
and $\Lambda(t) = \det(\mathsf{\Sigma})/m^2=a(t)c(t)-b(t)^2$. The functions
\begin{equation}
\label{equ:abc}
\begin{aligned}
a(t) &= [\dot G_+(t)]^2\Psi(t,t)
\\
b(t) &= \dot G_+(t)G_+(t)\left.
\frac{\partial \Psi(t,t')}{\partial t}\right\vert_{t'\nearrow t}
\\
c(t) &= [G_+(t)]^2\left.\frac{\partial^2 \Psi(t,t')}
{\partial t\partial t'}\right\vert_{t'\nearrow t}
\end{aligned}
\end{equation}
can all be expressed in terms of a single function
\begin{equation}
\label{equ:funcPsi}
\Psi(t,t') = \int_0^t {\rm d} s\int_0^{t'} {\rm d}u \,
K'(s-u) \frac{G_+(t-s)}{G_+(t)} \frac{G_+(t'-u)}{G_+(t')}\,.
\end{equation}
This function is completely determined by the thermal position autocorrelation
function $\langle q(t)q(0)\rangle$ and its time derivatives. Explicit
expressions are given in \ref{sec:AppendixB} and the interested reader may also
want to consult Ref.~\cite{GSI88} for further details.

The Gaussian form of the Wigner propagating function
(\ref{WignerPropExplSimpl}) is a consequence of the linearity of the harmonic
oscillator damped by the coupling to a harmonic heat bath. A similar
expression has therefore been found in the Markovian limit \cite{BO04}.
Similarly, as indicated in \ref{sec:AppendixA}, the result can be
generalized to the case of nonfactorizing initial conditions. Again one would 
find a Gaussian Wigner propagating function.

From (\ref{equ:DefFVWignerProp}) it follows that pairs of trajectories $q_\pm$
satisfying the equations of motion (\ref{equ:GenStationaryPathsHO}) and leading
to a sum trajectory connecting the initial phase-space point $\mathbf{r}'$ to
the end point $\mathbf{r}''$ contribute with a weight determined by
(\ref{equ:exponent2}). As discussed in the previous section, according to
(\ref{equ:GenStationaryPathsHO}) these pairs do not obey the classical damped
equation of motion. The same is true in phase space. The hyperbolic
character of the solutions of (\ref{equ:qxmotion}) already contains
all quantum effects that arise upon tracing out the heat bath. The
equations of motion (\ref{equ:GenStationaryPathsHO}) can therefore be
lifted to the phase space of the central oscillator by defining $p_\pm
= \dot q_\pm/m$,
\begin{equation}
\begin{aligned}
\dot p_\pm &= -m\omega_0^2q_\pm-m\gamma\dot q_\mp\\
\dot q_\pm &= \frac{p_\pm}{m}\,.
\end{aligned}
\end{equation}
so that the sum trajectory follows the classical equation of motion of the
damped harmonic oscillator. This result was also found by Ozorio and Brodier
\cite{ORB09} for the Markovian case derived directly from the
Lindblad master equation.

\begin{figure}
\centerline{\includegraphics[width=\columnwidth]{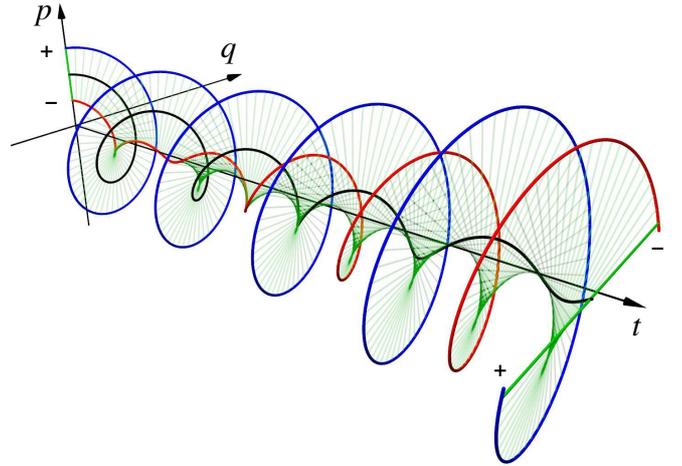}}
\caption{The time evolution of a pair of phase-space trajectories
$\mathbf{r}_\pm$ marked by $+$ (depicted in blue) and by $-$ (depicted in red)
is shown together with the corresponding classical sum trajectory depicted in
black. While the sum trajectory decays to zero for long times, the trajectories
$\mathbf{r}_\pm$ grow exponentially.}
\label{fig:trajectories}
\end{figure}

In Fig.~\ref{fig:trajectories} we depict the time evolution in phase space of
two trajectories $q_\pm$ indicated by $+$ and $-$ together with the
corresponding sum trajectory shown in black. Due to the damping, the sum
trajectory for long times approaches the origin of phase space. The
trajectories $q_\pm$ grow exponentially for long times and therefore clearly
behave nonclassically. Although in Fig.~\ref{fig:trajectories} the paths $q_+$
and $q_-$ have started on the same side of the origin of phase space, for long
times they are found opposite to each other. This is a consequence of their
exponential growth and of the fact that their midpoint approaches the origin of
phase space.

We close our discussion of the phase-space properties of the damped harmonic
oscillator by considering how the thermal equilibrium state is approached for
long times. First, we notice that in the Wigner propagator
(\ref{WignerPropExplSimpl}) the dependence on the initial phase-space
coordinates $\mathbf{r}'$ disappears in that limit because the sum coordinate
then approaches the origin of phase space. Furthermore, the long-time behavior
of the functions (\ref{equ:abc}) is given by
\begin{equation}
\label{equ:abcasymptot}
\begin{aligned}
a(t) &\sim \frac{m^2}{\hbar}
\frac{[\dot G_+(t)]^2}{[G_+(t)]^2}\langle q^2\rangle \\
b(t) &\sim -\frac{m^2}{\hbar}
\frac{[\dot G_+(t)]^2}{[G_+(t)]^2}\langle q^2\rangle \\
c(t) &\sim \frac{\langle p^2\rangle}{\hbar}+
\frac{m^2}{\hbar}\frac{[\dot G_+(t)]^2}{[G_+(t)]^2}\langle q^2\rangle\,,
\end{aligned}
\end{equation}
where $\langle q^2\rangle$ and  $\langle p^2\rangle$ are the second moments of
position and momentum, respectively, in thermal equilibrium. Inserting these
expressions into (\ref{WignerPropExplSimpl}), one obtains the thermal Wigner
function of the damped harmonic oscillator
\begin{equation}
\label{equ:thermalWignerFunc}
W_\beta(p,q) = \frac{1}{2\pi(\langle q^2\rangle\langle
p^2\rangle)^{1/2}}
\exp\left(-\frac{p^2}{2\langle p^2\rangle}-
\frac{q^2}{2\langle q^2\rangle}\right)\,.
\end{equation}

\begin{figure}
\centerline{\includegraphics[width=\columnwidth]{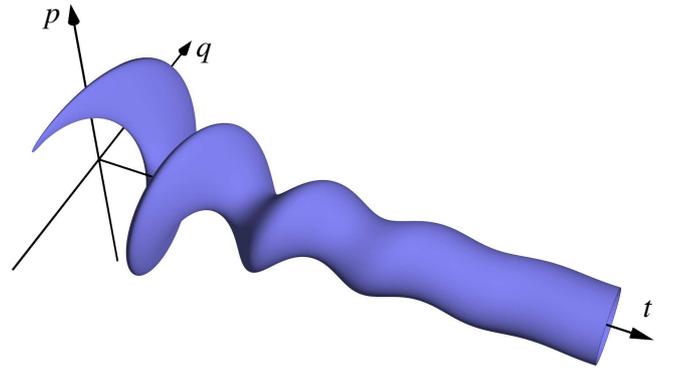}}
\caption{Isosurface of the time-dependent Wigner propagating function
(\ref{WignerPropExplSimpl}) for $\gamma/\omega_0=0.3$ and
$k_\mathrm{B}T=5\hbar\omega_0$. Position and momentum are scaled with respect
to the square root of the respective equilibrium second moments.}
\label{fig:Wigner3D}
\end{figure}

In Fig.~\ref{fig:Wigner3D} we illustrate the time evolution of the Wigner
propagating function for $\gamma=0.3\omega_0$ and
$k_\mathrm{B}T=5\hbar\omega_0$ by means of an isosurface. The function
(\ref{equ:funcPsi}) has been evaluated with the high-temperature approximation
$K'(t)=(2\gamma/\hbar\beta)\delta(t)$. The propagating function evolves from an
initial delta function to the thermal Wigner function
(\ref{equ:thermalWignerFunc}) which has a circular cross section because
position and momentum are scaled with the square roots of the respective 
equilibrium second moments.

%%%%%%%%%%%%%%%%%%%%%%%%%%%%%%%%%%%%%%%%%%%%%%%%%%%%%%%%%%%%%%%%%%%%%%%%%%%%%%
\section{Conclusion}
%%%%%%%%%%%%%%%%%%%%%%%%%%%%%%%%%%%%%%%%%%%%%%%%%%%%%%%%%%%%%%%%%%%%%%%%%%%%%%

Our analysis of the propagating function of a damped harmonic oscillator in
phase space has shown that pairs of trajectories contribute which, in contrast
to the undamped case, are nonclassical in nature. In the course of time the
distance between the two trajectories of a pair grows exponentially. This
behavior is typical also for nonlinear systems \cite{PDI10} so that the damped
harmonic oscillator can serve as an exactly solvable reference case.  The
harmonic oscillator coupled to a linear heat bath is however special insofar as
the linearity of the system implies that the sum of the pairs of
trajectories satisfies the classical damped equation of motion.

Furthermore, it was shown that it is the nonlocal interaction
(\ref{equ:exponent2}) appearing in a reduced description of the system degree
of freedom which is responsible for the broadening of the delta-like Wigner
propagator found for the undamped case into the Gaussian propagating function.
While most of the explicit results have been presented for the special case of
factorizing initial conditions, the generalization to a broader class of
initial states is straightforward along the lines presented in \ref{sec:AppendixA} 
and the Gaussian nature of the propagating function remains unchanged.

%%%%%%%%%%%%%%%%%%%%%%%%%%%%%%%%%%%%%%%%%%%%%%%%%%%%%%%%%%%%%%%%%%%%%%%%%%%%%%
\section*{Acknowledgements}
%%%%%%%%%%%%%%%%%%%%%%%%%%%%%%%%%%%%%%%%%%%%%%%%%%%%%%%%%%%%%%%%%%%%%%%%%%%%%%
The authors congratulate Peter H\"anggi to his 60$^\mathrm{th}$ birthday and
wish him many more healthful and productive years. One of us (GLI) would
particularly like to thank him on this occasion for a long-standing
collaboration and many fruitful discussions.

LAP is grateful for financial support by Colciencias and Universidad Nacional
de Colombia and the hospitality of the University of Augsburg.

%%%%%%%%%%%%%%%%%%%%%%%%%%%%%%%%%%%%%%%%%%%%%%%%%%%%%%%%%%%%%%%%%%%%%%%%%%%%%%
\appendix
%%%%%%%%%%%%%%%%%%%%%%%%%%%%%%%%%%%%%%%%%%%%%%%%%%%%%%%%%%%%%%%%%%%%%%%%%%%%%%
\section{Propagating function for nonfactorizing initial conditions}
\label{sec:AppendixA}
%%%%%%%%%%%%%%%%%%%%%%%%%%%%%%%%%%%%%%%%%%%%%%%%%%%%%%%%%%%%%%%%%%%%%%%%%%%%%%
In the main part of this paper, we have restricted our considerations to the
case of factorizing initial conditions for the sake of simplicity. The
generalization to nonfactorizing initial conditions is a bit more tedious but
straightforward.  In this appendix, we briefly derive the phase-space
representation of the time evolution of a nonfactorizing initial Wigner
function \cite{mer06}.

Specifically, we will consider nonfactorizing initial conditions where system
and bath together are in their equilibrium state thus accounting for initial
correlations. Then, operators taken from the Hilbert space of the system are
allowed to act in order to generate a nonthermal initial state for the system.
On the formal side, this preparation in comparison with factorizing initial
conditions has two consequences. Firstly, in addition to the two real-time
paths $q_\pm$, an imaginary-time path $\bar q$ appears. Secondly, a preparation
function $\lambda(\tilde{q}',q',\bar{\tilde{q}},\bar q)$ joins the
real-time and imaginary-time paths. Here, $\bar q$ and
$\bar{\tilde{q}}$ refer to sum and difference coordinates of the
imaginary time path. The preparation function function describes the
initial preparation and depends on the matrix elements of the system
operators. For more details, we refer the reader to Ref.~\cite{GSI88}.

The generalization of (\ref{rhoRPKRJ}) to the nonfactorizing initial conditions
just described reads
\begin{equation}
\label{equ:rhoSNonFact}
\begin{aligned}
\rho_{\rm S}(\tilde{q}'',q'',t) &=
\int\mathrm{d}\tilde{q}'\mathrm{d}q'\mathrm{d}\bar{\tilde{q}}
\mathrm{d}\bar q\,J(\tilde{q}'',q'',t;
\tilde{q}',q',{\bar{\tilde{q}}},{\bar q})\\
&\hphantom{=\int}\times\lambda(\tilde{q}',q',{\bar{\tilde{q}}},{\bar q})\,.
\end{aligned}
\end{equation}
Now, the propagating function also depends on the endpoints of the
imaginary-time path. Introducing the Wigner transform of the
preparation function
\begin{equation}
\begin{aligned}
\lambda_\mathrm{W}(p',q',\bar p,\bar q) &=
\frac{1}{(2\pi\hbar)^2}\int\mathrm{d}\tilde{q}'
\mathrm{d}\bar{\tilde{q}}\exp\left[\frac{\mathrm{i}}{\hbar}
(\bar p\bar{\tilde{q}}-p'\tilde{q}')\right]\\
&\hphantom{= \frac{1}{(2\pi\hbar)^2}\int}\times
\lambda(\tilde{q}',q',\bar{\tilde{q}},\bar q)\,,
\end{aligned}
\end{equation}
we obtain for the time evolution of the Wigner function after carrying out
the Fourier transform with respect to $\tilde{q}''$
\begin{equation}
\begin{aligned}
W_\mathrm{S}(p'',q'') &=
\int\mathrm{d}p'\mathrm{d}q'\mathrm{d}\bar p\mathrm{d}\bar q\,
G_\mathrm{W}(p'',q'',t;p',q',\bar p, \bar q)\\
&\hphantom{=\int}\times\lambda_\mathrm{W}(p',q',\bar p, \bar q)\,.
\end{aligned}
\end{equation}
By comparison with (\ref{equ:rhoSNonFact}) one finds for the relation
between the propagating function introduced in (\ref{equ:rhoSNonFact})
and its Wigner transform
\begin{equation}
\label{ArificialWignerProp}
\begin{aligned}
G_\mathrm{W}(p'',q'',t; p',q',{\bar p},{\bar q}) &=
\frac{1}{2\pi\hbar}
\int\mathrm{d}\tilde{q}''\mathrm{d}\tilde{q}'\mathrm{d}\bar{\tilde{q}}\\
&\hspace{-0.3 \columnwidth}
\times\exp\left[\frac{\mathrm{i}}{\hbar}
(p'\tilde{q}'-p''\tilde{q}''-\bar p\bar{\tilde{q}})\right]
J(\tilde{q}'',q'',t; \tilde{q}',q',\bar{\tilde{q}},\bar q)\,.
\end{aligned}
\end{equation}
For the special case of factorizing initial conditions, the
coordinates $\bar{\tilde{q}}$, $\bar q$ and the momentum $\bar q$ are
to be disregarded and one arrives at the relation
(\ref{equ:DefFVWignerProp}) between the propagating functions in
position and phase space.

%%%%%%%%%%%%%%%%%%%%%%%%%%%%%%%%%%%%%%%%%%%%%%%%%%%%%%%%%%%%%%%%%%%%%%%%%%%%%%
\section{Propagating function and position autocorrelation function}
\label{sec:AppendixB}
%%%%%%%%%%%%%%%%%%%%%%%%%%%%%%%%%%%%%%%%%%%%%%%%%%%%%%%%%%%%%%%%%%%%%%%%%%%%%%
The Gaussian nature of a harmonic oscillator coupled linearly to a bath of
harmonic oscillators implies that its reduced dynamics can be expressed
completely in terms of the thermal position autocorrelation function
\cite{GWT84,RHW85}
\begin{equation}
\begin{aligned}
C(t) &= \langle q(t)q(0)\rangle = S(t)+\mathrm{i}A(t)\\
&= \frac{\hbar}{\pi m}\int_{-\infty}^{+\infty}\mathrm{d}\omega
\frac{\omega\hat\gamma(-\mathrm{i}\omega)}
{(\omega^2-\omega_0^2)^2+\omega^2\hat\gamma(-\mathrm{i}\omega)^2}
\frac{\exp(-\mathrm{i}\omega t)}{1-\exp(-\beta\hbar\omega)}\,.
\end{aligned}
\end{equation}
$S(t)$ and $A(t)$ denote the symmetrized and antisymmetrized correlation
functions and correspond to the real and imaginary part of $C(t)$,
respectively. $\hat\gamma(z)$ is the Laplace transform of the friction kernel
(\ref{equ:frictionKernel}) divided by the oscillator mass $m$.  The
antisymmetric correlation function is related to the function $G_+(t)$
introduced for the special case of Ohmic damping in (\ref{equ:gpm}) by
\begin{equation}
G_+(t) = -\frac{2m}{\hbar}A(t)\Theta(t)\,,
\end{equation}
where $\Theta(t)$ is the unit step function. The second moments of position and
momentum appearing in (\ref{equ:thermalWignerFunc}) are related to the
symmetrized correlation function by $\langle q^2\rangle = S(0)$ and $\langle
p^2\rangle = -m\ddot S(0)$, respectively. For the latter to be finite, the
Laplace transform $\hat\gamma(z)$ requires a high-frequency cutoff.

The functions (\ref{equ:abc}) are found to read \cite{GSI88}
\begin{equation}
\begin{aligned}
a(t) &= \frac{m^2}{\hbar}\frac{[\dot G_+(t)]^2}{[G_+(t)]^2}
\left\{\langle q^2\rangle\left[1-\frac{[S(t)]^2}{\langle q^2\rangle^2}\right]
+\frac{\langle p^2\rangle}{m^2}[G_+(t)]^2+2\dot S(t)G_+(t)\right\}\\
b(t) &= -\frac{m^2}{\hbar}\frac{\dot G_+(t)}{G_+(t)}
\left\{\langle q^2\rangle\left[1-\frac{[S(t)]^2}{\langle q^2\rangle^2}\right]
\frac{\dot G_+(t)}{G_+(t)}+\dot G_+(t)\dot S(t)\right.\\
&\hphantom{= -\frac{m^2}{\hbar}\frac{\dot G_+(t)}{G_+(t)}\Big\{}
\left.-G_+(t)\ddot S(t)+\frac{S(t)\dot S(t)}{\langle q^2\rangle}\right\}\\
c(t) &= \frac{m^2}{\hbar}\left\{\langle q^2\rangle
\frac{[\dot G_+(t)]^2}{[G_+(t)]^2}+\frac{\langle p^2\rangle}{m^2}
-\frac{1}{\langle q^2\rangle}
\left[\dot S(t)-\frac{\dot G_+(t)}{G_+(t)}S(t)\right]^2\right\}\,.
\end{aligned}
\end{equation}
If one takes into account that $G_+(t)$, $S(t)$, and their derivatives decay to
zero for long times but not the ratio $\dot G_+(t)/G_+(t)$, one immediately
finds the asymptotic expressions (\ref{equ:abcasymptot}).


\section*{References}
\begin{thebibliography}{00}

\bibitem{vV28}
J. H. van Vleck, Proc. Natl. Acad. Sci. USA, 14 (1928) 178.

\bibitem{gut67}
M. C. Gutzwiller, J. Math. Phys. 8 (1967) 1979.

\bibitem{gut71}
M. C. Gutzwiller, J. Math. Phys. 12 (1971) 343.

\bibitem{HTB90}
P. H\"anggi, P. Talkner, M. Borkovec, Rev. Mod. Phys. 62 (1990) 251.

\bibitem{HH91}
P. H\"anggi, W. Hontscha, Ber. Bunsenges. Phys. Chem. 95 (1991) 379.

\bibitem{ber89}
M. V. Berry. Proc. R. Soc. Lond. A, 423 (1989) 219.

\bibitem{TAO01}
F. Toscano, M. A. M. de Aguiar, A. M. Ozorio
de Almeida, Phys. Rev. Lett. 86 (2001) 59.

\bibitem{DP09}
T. Dittrich, L. A. Pach\'on, Phys. Rev. Lett. 102 (2009) 150401.

\bibitem{RO02}
P. P. de M. Rios, A. M. Ozorio de Almeida, J. Phys. A: Math. Gen. 35
(2002) 2609.

\bibitem{DVS06}
T. Dittrich, C. Viviescas, L. Sandoval, Phys. Rev. Lett. 96 (2006) 070403.

\bibitem{DGP09}
T. Dittrich, E. A. G\'omez, L. A. Pach\'on, arXiv:0911.3871 (2009).

\bibitem{ORB09}
A. M. Ozorio de Almeida, P. de M. Rios, O. Brodier, J. Phys. A:
Math. Theor. 42 (2009) 065306.

\bibitem{PDI10}
L. A. Pach\'on, T. Dittrich, G.-L. Ingold, in preparation.

\bibitem{fey48}
R. P. Feynman, Rev. Mod. Phys. 20 (1948) 367.

\bibitem{FH65}
R. P. Feynman, A. R. Hibbs, Quantum mechanics and path integrals,
McGraw-Hill, New York, 1965.

\bibitem{HOSW84}
M. Hillery, R. F. O'Connell, M. O. Scully, E. P. Wigner,
Phys. Rep. 106 (1984) 121.

\bibitem{mar91}
M. S. Marinov, Phys. Lett. A 153 (1991) 5.

\bibitem{BV90} N. L. Balazs, A. Voros, Ann. Phys. (N. Y.) 199 (1990) 123.

\bibitem{U66}
P. Ullersma, Physica {\bf 32} (1966) 27.

\bibitem{PSZ08}
E. Pollak, J. Shao, D. H. Zhang, Phys. Rev. E 77 (2008) 021107.

\bibitem{FV63}
R. P. Feynman, F. L. Vernon Jr., Ann. Phys. (N. Y.) 24 (1963) 118.

\bibitem{CL83}
A. O. Caldeira,  A. J. Leggett, Physica A 121 (1983) 587.

\bibitem{wei93}
U. Weiss, Quantum Dissipative Systems, 	Series in Modern Condensed
Matter Physics, vol. 13,  World Scientific, Singapore, 2008.

\bibitem{HI05}
P. H\"anggi, G.-L. Ingold, Chaos 15 (2005) 026105.

\bibitem{HA85}
V. Hakim, V. Ambegaokar, Phys. Rev. A 32 (1985) 423.

\bibitem{GSI88}
H. Grabert, P. Schramm, G.-L. Ingold, Phys. Rep. 168 (1988) 115.

\bibitem{ZJH94}
C. Zerbe, P. Jung, P. H\"anggi, Phys. Rev. E 49 (1994) 3626.

\bibitem{ZH95}
C. Zerbe, P. H\"anggi, Phys. Rev. E 52 (1995) 1533.

\bibitem{KDH97}
S. Kohler, T. Dittrich, P. H\"anggi, Phys. Rev. E 55 (1997) 300.

\bibitem{HR85}
F. Haake, R. Reibold, Phys. Rev. A 32 (1985) 2462.

\bibitem{CYG02}
C.-I. Um, K.-H. Yeon, T. F. George, Phys. Rep. 362 (2002) 63.

\bibitem{BO04}
O. Brodier, A. M. Ozorio de Almeida, Phys. Rev. E 69 (2004) 016204.

\bibitem{mer06}
M. Merkl, Diploma thesis, Augsburg, 2006.

\bibitem{GWT84}
H. Grabert, U. Weiss, P. Talkner, Z. Phys. B 55 (1984) 87.

\bibitem{RHW85}
P. Riseborough, P. H\"anggi, U. Weiss, Phys. Rev. A 31 (1985) 471.

\end{thebibliography}
\end{document}